\documentclass{aa}
\usepackage{graphicx}
\begin{document}

\newcommand\beq{\begin{equation}}
\newcommand\eeq{\end{equation}}
\newcommand\mum{\mu{m}}
\def\DeltaRA{|\Delta{\rm RA|}}
\def\DeltaDec{|\Delta{\rm DEC|}}

{\title{Metallicity Estimates for Old Star Clusters in M33}}

\author{Jun Ma, Xu Zhou and Jiansheng Chen}
\offprints{Jun Ma, \\
\email{majun@vega.bac.pku.edu.cn}}
\institute{National Astronomical Observatories,
Chinese Academy of Sciences, Beijing, 100012, P.R. China}
\date{Received 18 February 2003 / Accepted 7 August 2003}

\abstract{ Using the theoretical stellar population synthesis
models of BC96, Kong et al. (2003) showed that some BATC colors
and color indices could be used to disentangle the age and
metallicity effect. They found that there is a very good relation
between the flux ratio of $L_{8510}/L_{9170}$ and the metallicity
for stellar populations older than 1 Gyr. In this paper, based on
the Kong et al. results and on the multicolor spectrophotometry of
Ma et al. (2001, 2002a,b,c), we estimate the metallicities of 31
old star clusters in the nearby spiral galaxy M33, 23 of which are
``true'' globular clusters. The results show that most of these
old clusters are metal poor. We also find that the ages and metal
abundance for these old star clusters of M33 do not vary with
deprojected radial position.
}

\titlerunning{Metallicity Estimates for Old Star Clusters in M33}
\authorrunning{J. Ma et al.}
\maketitle

\section{Introduction}

Globular clusters (GCs) are thought to be among the oldest radiant
objects in the Universe. They are simple coeval stellar systems
which formed on a very short timescale during phases of intense
star formation in host galaxies. The GCs of the Milky Way
probe the manner in which our Galaxy formed. Studies of similar
populations in other galaxies can reveal the properties of
these galaxies soon after their formation. For example, the widely
varying specific frequency of GCs in individual galaxies indicates
that cluster formation is almost certainly affected by the local
environment with the host galaxy.

M33, at a distance of 850 kpc, is the third-brightest member of
the Local Group, and is classified as a late-type ScII-III spiral
(\cite{van99}). This galaxy represents a morphological type
intermediate between the largest ``early-type'' spirals and the
dwarf irregulars in the Local Group (Chandar, Bianchi, \& Ford
1999a). M33 subtends $1^{\circ}$ on the sky. Its large angular
extent and favorable inclination $i=56^{\circ}$ (Regan \& Vogel
1994) make it suitable for studies of stellar content. However,
its large size makes its study difficult. The
Beijing-Arizona-Taiwan-Connecticut (BATC) Multicolor Sky Survey
(Fan et al. 1997; Zheng et al. 1999) obseves M33 as part of its
galaxy calibration program.

Before Chandar et al. (1999a, 1999b, 1999c, 2001, 2002), the
system of clusters in M33 were not be well studied, although
detection, and photometry and spectrophotometry have been obtained
(see details from Hiltner 1960; Kron \& Mayall 1960; Melnick \&
D'Odorico 1978; Christian \& Schommer 1982, 1988; Huchra et al.
1996). Now, a database of about 400 star clusters is available
from the ground-based work, and from the {\it {Hubble Space
Telescope}} (HST) images.

Using the theoretical stellar population synthesis models of
Bruzual \& Charlot (1996, unpublished, hereafter BC96) and
multicolor photometry, Kong et al. (2000) studied the age,
metallicity, and interstellar-medium reddening distribution for
M81. When they convolved the spectral energy distributions (SEDs)
of BC96 with the BATC filter profiles to obtain the optical and
near-integrated luminosity, they found that, among all the BATC
filter bands, the color index centered at 8510{\AA} is much more
sensitive to the metallicity than to the age. The center of this
filter band is near the Ca~II triplets ($\lambda\lambda=8498,
8542, 8662${\AA}) (hereafter CaT). As Zhou (1991) noted, that the
strength of the CaT depends on the effective temperature, surface
gravity, and the metallicity in late-type stars. A very good
relation between the flux ratio of $I_{8510}\equiv
L_{8510}/L_{9170}$ and the metallicity was found for stellar
populations older than 1 Gyr in Kong et al. (2000).

In this paper, we estimate the metallicities of the 31 old star
clusters that were detected by Christian \& Schommer (1982),
Chandar, Bianchi, \& Ford (1999a, 2001), and by Mochejska et al.
(1998) in M33 using the relation of Kong et al.
(2000). The ages of all these sample star clusters
were estimated by Ma et al. (2001, 2002a, 2002b, 2002c) by comparing the
BC96 simple stellar population synthesis models with
the integrated photometric measurements.

The outline of the paper is as follows. Sample selection,
observations and data reduction are given in section 2. Section 3
presents the spectral synthesis models. In section 4, we provide a
brief description of the method of Kong et al. (2000), and
estimate the metallicities of 31 sample old star
clusters. Some statistical properties of old star clusters are
investigated in section 5. In section 6, we give our major results
and some discussions.

\section{Sample of star clusters, observations and data reduction}

\subsection{Sample of old star clusters}

The sample of star clusters in this paper is from Ma et al. (2001,
2002a,b,c), who presented multicolor photometry and estimated the
ages using BC96 models for 180 star clusters in M33. The RA and
Dec of these clusters are from Christian \& Schommer (1982),
Chandar, Bianchi, \& Ford (1999a, 2001), or Mochejska et al.
(1998). Christian \& Schommer (1982) detected more than 250
nonstellar objects using $14\times 14$ $\rm inch^2$ unfiltered,
unbaked, IIa-O focus plates exposed for 150 minutes with the Kitt
Peak 4 m Richey-Chr\'{e}tien (R-C) direct camera. Chandar, Bianchi
\& Ford (1999a, 2001) used 55 multiband HST WFPC2 fields to search
for star clusters much closer to the nucleus of M33 than previous
studies, and detected 162 star clusters, 131 of which were
previously unknown. Mochejska et al. (1998) detected 51 globular
cluster candidates in M33, 32 of which were not previously
cataloged, using the data collected in the DIRECT project
({\cite{Kaluzny98}; {\cite{Stanek98}). Ma et al. (2001, 2002a,b,c)
obtained the SEDs of the 180 clusters by aperture photometry, and
estimated their ages using the theoretical evolutionary population
synthesis methods. In Ma et al. (2001), there are 10 clusters, the
ages of which are older than 1 Gyr. We exclude three star clusters
of these 10 because of their low signal-to-noise ratios. Also,
cluster 54 in Chandar, Bianchi, \& Ford (1999a) is U137 in
Christian \& Schommer (1982). In Ma et al. (2002b), there are 22
clusters older than 1 Gyr. However, the ratios of signal-to-noise
of 11 clusters are not high enough, and are not included in this
sample. In Ma et al. (2002c), there are 5 clusters older than 1
Gyr. The signal-to-noise ratio of cluster 2 is not high enough,
and is also not included in this sample. Altogether, there are 31
old star clusters in this paper. Figure 1 is the image of M33 in
filter BATC07 (5785{\AA}), the circles indicate the positions of
the sample clusters. By comparing the photometric measurements to
integrated colors from theoretical models by Bertelli et al.
(1994), Chandar et al. (1999b, 2002) estimated ages for 23 old
star clusters in common. Table 1 lists the comparison of age
estimates with previously results (Chandar et al. 1999b, 2002).
Except for clusters 49 and 59 of Chandar et al. (1999a), the ages
estimated by Ma et al. (2001, 2002a,b,c) are consistent with the
ones estimated by Chandar et al. (1999b, 2002). Cluster 49, the
$B-V$ value of which is very large $0.824$ (Chandar et al. 1999b),
should be an old cluster. Our sample includes 23 ``true'' globular
clusters, which have $(B-V)_0\geq 0.6$ or $(V-I)_0\geq 0.78$,
colors typical of Galactic GCs (Chandar, Bianchi, \& Ford 2001).

\begin{figure}
\vspace{6.5cm}
\caption{The image of M33 in filter BATC07 (5785{\AA}) and the positions
of the sample star clusters.
The center of the image is located at
${\rm RA=01^h33^m50^s{\mbox{}\hspace{-0.13cm}.}58}$
Dec=30$^\circ39^{\prime}08^{\prime\prime}{\mbox{}\hspace{-0.15cm}.4}$
(J2000.0). North is up and east is to the left.}
\label{fig1}
\end{figure}

\subsection{Observations and data reduction}

The large field multicolor observations of the spiral galaxy M33
were collected using the Ford Aerospace $2048\times 2048$ CCD
mosaic camera on the 60/90 cm f/3 Schmidt telescope of the
Xinglong station of the National Astronomical Observatories. The
field of view of the CCD is $58^{\prime}$ $\times $ $ 58^{\prime}$
with a pixel scale of $1\arcsec{\mbox{}\hspace{-0.15cm}.} 7$. The
typical seeing of the Xinglong station is $2\arcsec$. The
multicolor BATC filter system, which was specifically designed to
avoid contamination from the brightest and most variable night sky
emission lines, includes 15 intermediate-band filters, covering
the total optical wavelength range from 3000 to 10000{\AA}. It is
defined the magnitude zero points similar to the
spectrophotometric AB magnitude system, a $\tilde{f_{\nu}}$
monochromatic system (Oke \& Gunn 1983) based on the SEDs of the
four $F$ sub-dwarfs, HD~19445, HD~84937, BD~${+26^{\circ}2606}$,
and BD~${+17^{\circ}4708}$. The advantage of the AB magnitude
system is that the magnitude is directly related to physical
units. The BATC magnitude system is defined as the AB magnitude
system,
\begin{equation}
m_{\rm batc}=-2.5{\rm log}\tilde{F_{\nu}}-48.60,
\end{equation}
where $\tilde{F_{\nu}}$ is the appropriately averaged monochromatic
flux in units of erg s$^{-1}$ cm$^{-2}$ Hz$^{-1}$ at
the effective wavelength of the specific passband.
In the BATC system (Yan et al. 1999), $\tilde{F_{\nu}}$ is
defined as
\begin{equation}
\tilde{F_{\nu}}=\frac{\int{d} ({\rm
log}\nu)f_{\nu}R_{\nu}} {\int{d} ({\rm log}\nu)R_{\nu}},
\end{equation}
which links the magnitude to the number of photons detected by the
CCD rather than to the input flux (Fukugita et al. 1996). In this
equation, $R_{\nu}$ is the system response, $f_{\nu}$ is the
SED of the source.

The images of M33 covering the whole optical body of M33 were
accumulated in 13 intermediate band filters with a total exposure
time of about 37.25 hours from September 23, 1995 to August 28,
2000. Data reduction, by bias subtraction and flat-fielding with
dome flats, was performed with the automatic data reduction
software PIPELINE I developed for the BATC multicolor sky survey
(\cite{Fan96}; \cite{Zheng99}). The dome flat-field images were
taken by using a diffuse plate in front of the correcting plate of
the Schmidt telescope. We performed photometric calibration of the
M33 images using the Oke-Gunn primary flux standard stars
HD~19445, HD~84937, BD~${+26^{\circ}2606}$, and
BD~${+17^{\circ}4708}$, which were observed during photometric
nights (see details from Yan et al. 1999; Zhou et al. 2001).

Using the images of the standard stars observed on photometric
nights, we derive iteratively the extinction curves and the slight
variation of the extinction coefficients with time (Zhou et al.
2001). The extinction coefficients at any given time in a night $[K+
\Delta K (UT)]$ and the zero point of the instrumental magnitude
($C$) were obtained by

\begin{equation}
m_{\rm batc}=m_{\rm inst}+[K+\Delta K(UT)]X+C,
\end{equation}
where $X$ is air mass. The instrumental magnitudes ($m_{\rm
inst}$) of the selected bright, isolated and unsaturated stars on
the M33 images of the same photometric nights can be readily
transformed to the BATC AB magnitude system ($m_{\rm batc}$). We
calibrated the photometry on the combined images by comparing the
magnitudes of these stars to determine a mean magnitude offset to
the photometric images. Table 2 lists the parameters of the BATC
filters and the statistics of observations. Column 6 of Table 2
gives the zero point error, in magnitude, for the standard stars
in each filter. The formal errors we obtain for these stars in the
13 BATC filters are $\la 0.02$ mag. This indicates that we can
define the standard BATC system to an accuracy of  $\la 0.02$ mag.

\subsection{Integrated photometry}
For each star cluster, aperture photometry was used to obtain
magnitudes. To avoid contamination from nearby objects, we adopt a
small aperture of $6\arcsec{\mbox{}\hspace{-0.15cm}.} 8$
corresponding to a diameter of 4 pixels in the Ford CCD.} The
large-aperture measures on the uncrowded bright stars were used to
determine the aperture corrections, i.e., the magnitude difference
between the small-aperture magnitude and the ``total'' or
seeing-independent magnitude for the stars on each frame. The
uncertainties for each filter take into account the error from the
object count rate, sky variance, and instrument gain. For
convenience, we present Table 3 with multiband photometry for
sample clusters in this paper based on Ma et al. (2001,
2002a,b,c). Column 1 is cluster number. Column 2 to Column 14 show
the magnitudes of different bands. The second line of each star
cluster is the uncertainties of magnitude of corresponding band.

\section{Spectral synthesis}
Since the pioneering works of Tinsley (1972) and Searle, Sargent,
\& Bagnuolo (1973), spectral population synthesis has become a
standard technique to study the stellar populations of galaxies.
Since stellar clusters can be assumed single age and single
metallicity group of stars, their integrated colors reflect their
age and metallicity for a given initial mass function. A
comprehensive compilation of various evolutionary synthesis models
was presented by Leitherer et al. (1996) and Kennicutt (1998). One
of the widely used models is BC96. In this model, the evolution of
the spectrophotometric properties for a wide range of stellar
metallicity, $Z=0.0004, 0.004, 0.008, 0.02, 0.05,$ and $0.1$, are
presented. The evolving spectra include the contribution of the
stellar component in the range from the EUV to the FIR. The age
varies from 0 to 20 Gyr and various IMFs are considered.

To estimate the ages of star clusters in M33, we convolve the SED
of BC96 with BATC filter profiles to obtain the optical and
near-infrared integrated luminosity. The integrated luminosity
$L_{\lambda_i}(t,Z)$ of the $i$th BATC filter can be calculated as
\beq L_{\lambda_i}(t,Z) =\frac{\int
F_{\lambda}(t,Z)\varphi_i(\lambda)d\lambda} {\int
\varphi_i(\lambda)d\lambda}, \eeq where $F_{\lambda}(t,Z)$ is the
SED of the BC96 of metallicity $Z$ at age $t$,
$\varphi_i(\lambda)$ is the response functions of the $i$th filter
of the BATC filter system ($i=3, 4, \cdot\cdot\cdot, 15$),
respectively. To avoid using distance-dependent parameters, we
calculate the integrated colors of a BC96 relative to the BATC
filter BATC08 ($\lambda=6075${\AA}):

\beq \label{color}
C_{\lambda_i}(t,Z)={L_{\lambda_i}(t,Z)}/{L_{6075}(t,Z)}.
\eeq

As a result, we obtain intermediate-band colors for 6
metallicities from $Z=0.0004$ to $Z=0.1$. Then, we determined the
ages and best-fit models of metallicity by minimizing the
difference between the intrinsic and integrated colors of BC96,

\beq R^2(n,t,Z)=\sum_{i=3}^{15}[C_{\lambda_i}^{\rm
intr}(n)-C_{\lambda_i}^ {\rm ssp}(t, Z)]^2,
\eeq
where$C_{\lambda_i}^{\rm ssp}(t, Z)$ represents the integrated
color in the $i$th filter of a SSP with age $t$ and metallicity
$Z$, and $C_{\lambda_i}^{\rm intr}(n)$ is the intrinsic integrated
color for nth star cluster. For convenience, we also list the ages
of sample star clusters of this paper in Column 5 of Table 4. The
uncertainties in the age estimates arising from photometric
uncertainties are 0.2 or so, i.e, $\rm{age}\pm
0.2\times\rm{age}~[\log \rm{yr}]$, and are formal errors that do
not include the model uncertainties. The star cluster ages
obtained in this paper are model-dependent and do not represent
``absolute values''. Uncertainties exist in the stellar evolution,
in the physics of the stellar structure and in the spectral
libraries. For example, Charlot, Worthey, \& Bressan (1996)
evaluated the uncertainties in stellar population synthesis models
by analyzing in detail the origin of the discrepancies between
three models (Bertelli et al. 1994; Worthey 1994; BC96), and
showed the main uncertainties originate from the underlying
stellar evolution theory, the color-temperature scale of giant
stars, and the flux libraries. Cardiel et al. (2003) also
discussed in detail the problem of disentangling stellar
population properties using the spectroscopic data. Vazdekis et
al. (2001) investigated the origin of the discrepancy between the
spectroscopic age and the CMD age for the Milky Way GC 47 Tuc, and
found that the ${\alpha}$-enhanced isochrones with atomic
diffusion included can provide a good fit to the CMD of 47 Tuc and
lead to a spectroscopic age in better agreement with the CMD age.

\section{Metallicity estimates}

\subsection{Correlation between color index and metallicity}

To study the integrated properties of the stellar population in
M81, Kong et al. (2000) used the simple stellar population
synthesis models of BC96. First, they convolved the SEDs of BC96
with the BATC filter profiles to obtain the optical and
near-infrared integrated luminosity. When they plot the relations
between color and age, they found that, among all the BATC filter
bands, the color index centered at 8510{\AA} is much more
sensitive to the metallicity than to the age (see Fig. 3 of Kong
et al. 2000 for details). The center of this filter band
(8510{\AA}) is near the CaT. A good relationship between the flux
ratio of $I_{8510}\equiv L_{8510}/L_{9170}$ and the metallicity
for stellar populations older than 1 Gyr was found (Eq. (4) of
Kong et al. 2000),

\beq
Z=(0.83-0.84\times I_{8510})^2.
\eeq

\subsection{Results}

Using equation (7), we can calculate the metallicities of these
old star clusters. We obtained $I_{8510}\equiv L_{8510}/L_{9170}$
using the photometric magnitudes in BATC13 and BATC14 bands (see
Table 3). Then, the metallicities can be obtained using equation
(7). The results are listed in Table 4 ($[\rm {Fe/H}] =\log Z-\log
Z_{\odot}$). The uncertainties for the metallicities are just the
formal errors, since Kong et al. (2000) did not discuss any errors
and uncertainties when they derived Eq. (7). The formal errors are
obtained in the following way. Random values are selected for the
observed data such that they obey a normal distribution, with
sigma determined by the known errors in each sampled bin. We then
obtain the best-fit metallicity. This procedure was repeated 300
times, giving us 300 separate determinations of the best-fit
metallicity. The statistical standard deviation of metallicity
from this procedure is adopted as the final error for the
metallicity. In Table 4, we also present the ages for the sample
clusters from Ma et al. (2001, 2002a,b,c). Table 5 lists some
sample cluster metallicities from other authors (Cohen, Persson,
\& Searle 1984; Christian \& Schommer 1988; Brodie \& Huchra 1991;
Sarajedini et al. 1998). Using two reddening-independent
techniques, Cohen, Persson, \& Searle (1984) obtained abundance
estimates for the four GCs, M9, U49, H38, and C20. Comparing the
results for these four clusters, we find that the only very
discrepant result is for M9, which we derive to be moderately
metal rich, while Cohen, Persson, \& Searle (1984) estimated it to
be very metal poor. However, the results of Christian \& Schommer
(1988) and Sarajedini et al. (1998) for M9 are intermediate
between Cohen, Persson, \& Searle (1984) and this study. Christian
\& Schommer (1988) and Brodie \& Huchra (1991) estimated the
metallicities for the 10 GCs using integrated spectra. The mean
metallicity difference (the values of this paper minus the values
of Christian \& Schommer 1988 and Brodie \& Huchra 1991) is
$<\Delta [\rm {Fe/H}]>=0.153\pm 0.192$. Sarajedini et al. (1998)
estimated the metallicities for these 10 GCs based on the shape
and color of the red giant branch. Our results are consistent with
Sarajedini et al. (1998) except for C38, which we find is very
metal poor, but Sarajedini et al. (1998) find it to be the most
metal rich. The mean metallicity difference (the values of this
paper minus the values of Sarajedini et al. 1998) is $<\Delta [\rm
{Fe/H}]>=0.148\pm 0.216$. Sarajedini et al. (2000) estimated the
cluster metallicities using the integrated $B-V$ colors from
Christian \& Schommer (1988) and the equations of Couture, Harris,
\& Allwright (1990), and presented that the metallicity of R12 is
very metal rich. Except for this cluster, our results are also
consistent with the ones obtained using $B-V$ colors, and the mean
difference ($<{\rm[Fe/H]_{This~paper}-[Fe/H]}_{B-V}>$) is
$0.056\pm 0.273$.

\section{Some properties of old star clusters}

The following statistical relations are based on our data and
are thus model-dependent.

\subsection{Metallicity as a function of deprojected distance}

V\'\i lchez et al. (1988) studied the abundance gradient in M33 on
the data of emission lines in selected HII regions. The O/H
gradient is steep in the inner regions, but much flatter in the
outer regions, and N/O is constant over most of the visible disk,
but lower in the outer HII region. We can investigate the radial
abundance behavior of the old clusters in M33. Using the data of
both the red and blue portions of the instability trip of two halo
globular clusters (M9 and U77) and the RR Lyrae luminosity
relation, Sarajedini et al. (2000) estimated M33 to be at a
distance of $(m-M)_0=24.84\pm 0.16$, which is adopted in this
paper. We also adopted the inclination and position angles to be
$56^{\circ}$ and $23^{\circ}$ of M33, respectively (Regan \& Vogel
1994). When the line of intersection (i.e. the major axis of the
image) between the galactic plane and tangent plane is taken as
the polar axis, it is easily proved that

\begin{equation}
r=\rho\sqrt{1+\tan^2 \gamma\sin^2\theta}
\end{equation}
and
\begin{equation}
\tan\phi=\frac{\tan\theta}{\cos\gamma},
\end{equation}
where $r$ and $\phi$ are the polar co-ordinates in the galactic
plane, and $\rho$ and  $\theta$ are the corresponding co-ordinates
in the tangent plane, and $\gamma$ is the inclination angle of the
galactic disk. Using formula (8), we can obtained the
distances of our sample clusters from the center of M33. Figure 2
displays the variation of metal abundance with deprojected radial
position in units of kpc for 23 globular cluster candidates in
M33. This figure presents no relationship between the metal
abundance of a cluster and its distance from the galactic center.
This conclusion is consistent with that in Sarajedini et al.
(2000) for the 9 GCs in this galaxy. Figure 3 plots
the variation of metal abundance with deprojected radial position
for all old clusters in this study.

\begin{figure}
\resizebox{\hsize}{!}{\rotatebox{-90}{\includegraphics{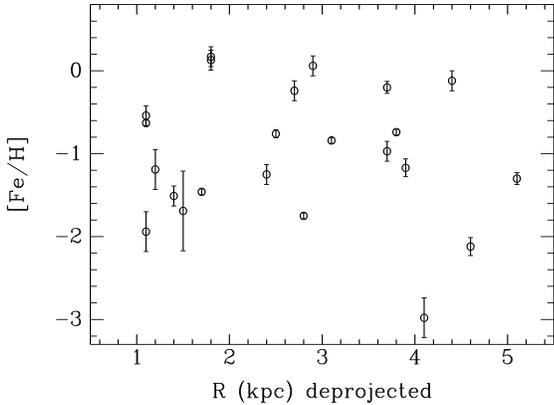}}}
\vspace{-1cm} \caption{Deprojected radial variation of metal
abundance for 23 ``true'' GCs of M33.}
\label{fig2}
\end{figure}

\begin{figure}
\resizebox{\hsize}{!}{\rotatebox{-90}{\includegraphics{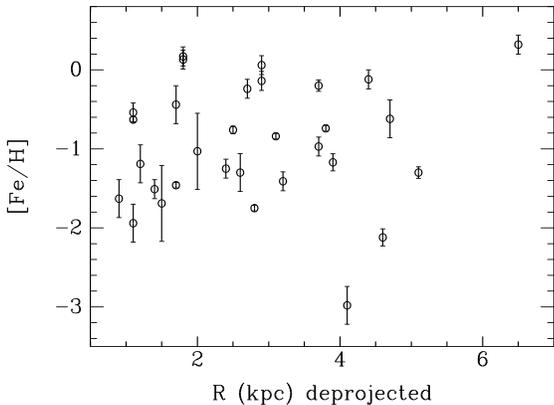}}}
\vspace{-1cm} \caption{Deprojected radial variation of metal
abundance for 31 old clusters of M33.}
\label{fig3}
\end{figure}

\subsection{Age as a function of deprojected distance}

In Figure 4, we plot the relation between age and
galactocentric distance for 23 ``true'' GCs of M33.
Here we note that no relationship exists. This conclusion is
consistent with Sarajedini \& King (1989) and
Chaboyer, Demarque, \& Sarajedini (1996) for the GCs
in the Galaxy. Figure 5 shows this relation for all 31 old
clusters in this study, and no relationship can be found, too.

\begin{figure}
\resizebox{\hsize}{!}{\rotatebox{-90}{\includegraphics{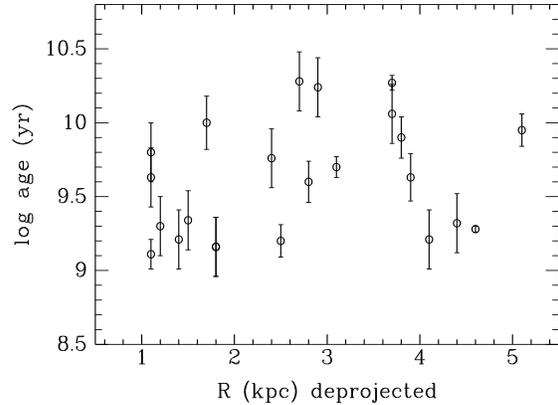}}}
\vspace{-1cm} \caption{Age as a function of galactocentric
distance for 23 ``true'' GCs of M33.}
\label{fig4}
\end{figure}

\begin{figure}
\resizebox{\hsize}{!}{\rotatebox{-90}{\includegraphics{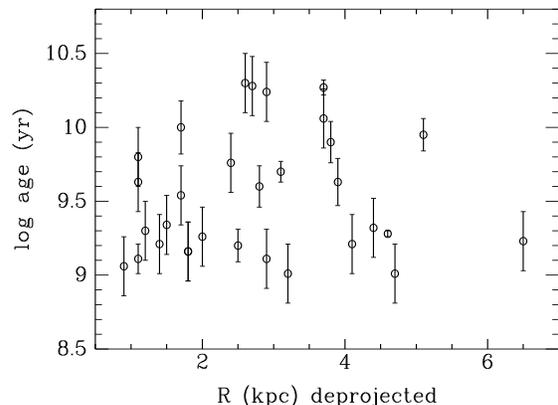}}}
\vspace{-1cm} \caption{Age as a function of galactocentric
distance for 31 old clusters of M33.}
\label{fig5}
\end{figure}

\section{Summary and discussion}

In this paper, based on the results of multicolor
spectrophotometry in Ma et al. (2001, 2002a, 2001b, 2002c) and on
the formula of Kong et al. (2000), we estimate the metallicities
of 31 old star clusters in M33, of which there are 23 ``true''
GCs. The results show that most of these old clusters are metal
poor. At the same time, we compare our results with others (Cohen,
Persson, \& Searle 1984; Christian \& Schommer 1988; Brodie \&
Huchra 1991; Sarajedini et al. 1998) derived using different
methods, such as integrated spectra and photometry. In general,
our results are consistent. The statistical results show that, the
ages and metal abundances based on our data do not vary with
deprojected radial position.

As we know, the old stellar populations and the nuclei of spiral
galaxies are dominated by G, K and M stars and therefore emit bulk
of their light in the near infrared region of the spectrum.
The CaT feature has been the subject of several analyses. Different
authors emphasized its different utilizations, such as a
luminosity indicator, and a possible discriminator between the
light contribution due to dwarfs and giants in a given population
mix (\cite{Idiart97}). Bica \& Alloin (1987) presented CCD spectra
with 12.5{\AA} resolution for 30 star clusters, and measured the
near-infrared continuum distribution and the equivalent widths of
13 absorption features. They found that in the
near-infrared spectral range, metallicity is the dominant
parameter. Based on the analysis of stars (giant stars,
supergiants and dwarfs), star clusters, and galaxy nuclei, Alloin
\& Bica (1989) showed that the equivalent widths of CaT are
metallicity dependent, although not as much as other metallic or
molecular features (CN or Mg+MgH). Armandroff \& Zinn (1988) found
that the strength of the CaT in the integrated spectra of GCs
forms a one-parameter family, this parameter being the
metallicity. Erdelyi-Mendes \& Barbuy (1991) calculated synthetic
spectra for the Ca~II lines in the local thermodynamic equilibrium
approximation using the model atmospheres computed by
interpolation in the grids of models by Gustafsson et al. (1975)
and an unpublished grid of dwarf models (see Erdelyi-Mendes \&
Barbuy 1991 for details). By a detailed analysis of the behavior
of the strength of Ca~II lines as a function of stellar
parameters, Erdelyi-Mendes \& Barbuy (1991) concluded that CaT has
a weak dependence on the effective temperature, a modest
dependence on surface gravity, but a quite important dependence on
metallicity, i.e., an exponential dependence between the flux of
the 2 strongest Ca~II lines ($\lambda\lambda$ 8542{\AA} and
8662{\AA}) and metallicity. Mallik (1994) presented the
observations of the infrared triplet lines of ionized Ca for 91
stars in the spectral range $\rm{F8-M4}$ of all luminosity classes
and in the metallic range $-0.65-+0.60$, and
found that the dependence of the CaT fluxes on gravity and
metallicity is intricately connected, but for supergiants a
strong relationship can be found. He also indicated
that when a large metallicity range is considered (i.e. $\geq
1.0$ dex), the influence of the metallicity on the CaT becomes
conspicuous. Idiart et al. (1997) also confirmed the strong
dependence of CaT index on metallicity.

\begin{acknowledgements}
We would like to thank the anonymous referee for his/her
insightful comments and suggestions that improved this paper. This
work has been supported by the National Key Basic Research Science
Foundation (NKBRSF TG199075402) and in part by the National
Science Foundation
\end{acknowledgements}


\clearpage
\setcounter{table}{0}
\begin{table}[htb]
\caption[]{Comparion age estimates with previous measurements}
\vspace {0.3cm}
\begin{tabular}{ccc}
\hline
\hline
                            &  Chandar et al. & Ma et al. \\
$\rm {Cluster}^{a}$ & $\log$ age (yr) & $\log$ age (yr) \\
\hline
         U49 &  9.2 $\pm$    0.1 &   9.60\\
         R12 &  9.7 $\pm$    0.1 &  10.00\\
         R14 & 10.2 $\pm$    0.2 &   9.11\\
          M9 &  9.2 $\pm$    0.1 &   9.63\\
         U77 & 9.15 $\pm$   0.15 &   9.20\\
         H38 & 9.25 $\pm$   0.15 &   9.70\\
         C20 &  9.2 $\pm$    0.1 &   9.95\\
         C38 &  8.9 $\pm$    0.1 &   9.28\\
         H10 & 9.25 $\pm$   0.15 &   9.90\\
        U137 & 9.35 $\pm$   0.15 &  10.27\\
       CBF11 & 9.50 $\pm$   0.30 &  10.30\\
       CBF20 & 10.1 $\pm$   0.20 &   9.54\\
       CBF22 & 9.25 $\pm$   0.15 &   9.26\\
       CBF28 & 10.2 $\pm$   0.40 &   9.80\\
       CBF49 & 7.90 $\pm$   0.20 &   9.34\\
       CBF59 & 7.40 $\pm$   0.20 &   9.11\\
       CBF69 &  9.2 $\pm$    0.1 &   9.76\\
       CBF74 &  9.3 $\pm$    0.1 &   9.32\\
       CBF97 &  9.3 $\pm$    0.2 &  10.28\\
      CBF112 &  8.8 $\pm$    0.2 &   9.21\\
      CBF118 & 9.15 $\pm$   0.15 &   9.16\\
      CBF161 &  9.1 $\pm$    0.1 &   9.23\\
         M12 &  9.4 $\pm$    0.2 &   9.63\\
\hline
\end{tabular}\\
\end{table}
\vspace{-0.5cm}
$^{\rm a}~${CBF identifications are from
Chandar, Bianchi, \& Ford (1999a, 2001);
M identifications are from Mochejska et al. (1998);
The others are from Christian \& Schommer (1982).}

\clearpage
\setcounter{table}{1}
\begin{table}[htb]
\caption[]{Parameters of the BATC Filters and Statistics of Observations}
\vspace {0.5cm}
\begin{tabular}{cccccc}
\hline
\hline
 No. & Name& cw$^{\rm a}$(\AA)& Exp. (hr)&  N.img$^{\rm b}$
 & rms$^{\rm c}$ \\
\hline
1  & BATC03& 4210   & 00:55& 04 &0.024\\
2  & BATC04& 4546   & 01:05& 04 &0.023\\
3  & BATC05& 4872   & 03:55& 19 &0.017\\
4  & BATC06& 5250   & 03:19& 15 &0.006\\
5  & BATC07& 5785   & 04:38& 17 &0.011\\
6  & BATC08& 6075   & 01:26& 08 &0.016\\
7  & BATC09& 6710   & 01:09& 08 &0.006\\
8  & BATC10& 7010   & 01:41& 08 &0.005\\
9  & BATC11& 7530   & 02:07& 10 &0.017\\
10 & BATC12& 8000   & 03:00& 11 &0.003\\
11 & BATC13& 8510   & 03:15& 11 &0.005\\
12 & BATC14& 9170   & 04:45& 25 &0.011\\
13 & BATC15& 9720   & 05:00& 26 &0.009\\
\hline
\vspace{0.1cm}
\end{tabular}\\
$^{\rm a}$ Central wavelength for each BATC filter\\
$^{\rm b}$ Image numbers for each BATC filter\\
$^{\rm c}$ Zero point error, in magnitude, for each filter
as obtained\\
from the standard stars
\end{table}

\clearpage
\setcounter{table}{2}
\begin{table}[ht]
\caption{SEDs of 31 Old Star Clusters}
\vspace {0.3cm}
\begin{tabular}{cccccccccccccc}
\hline
\hline
$\rm {Cluster}^{a}$  & 03  &  04 &  05 &  06 &  07 &  08 &  09 &  10 &  11 &  12 &  13 &  14 &  15\\
(1)    & (2) & (3) & (4) & (5) & (6) & (7) & (8) & (9) & (10) & (11) & (12) & (13) & (14)\\
\hline
     U49 &  16.99 &  16.56 &  16.46 &  16.27 &  16.06 &  15.98 &  15.82 &  15.78 &  15.70 &  15.58 &  15.52 &  15.48 &  15.40\\
 & 0.022 &  0.018 &  0.014 &  0.015 &  0.010 &  0.011 &  0.011 &  0.010 &  0.010 &  0.009 &  0.010 &  0.012 &  0.012    \\
     R12 &  17.34 &  16.81 &  16.62 &  16.43 &  16.15 &  16.11 &  15.90 &  15.83 &  15.77 &  15.68 &  15.56 &  15.51 &  15.44\\
&  0.049 &  0.036 &  0.027 &  0.029 &  0.020 &  0.021 &  0.022 &  0.023 &  0.022 &  0.019 &  0.021 &  0.020 &  0.024     \\
     R14 &  17.55 &  17.04 &  16.84 &  16.59 &  16.21 &  16.14 &  15.91 &  15.81 &  15.66 &  15.55 &  15.43 &  15.32 &  15.20\\
&  0.052 &  0.037 &  0.027 &  0.028 &  0.019 &  0.021 &  0.017 &  0.018 &  0.020 &  0.016 &  0.017 &  0.017 &  0.021     \\
     M9  &  17.82 &  17.48 &  17.38 &  17.17 &  16.96 &  16.90 &  16.73 &  16.68 &  16.60 &  16.53 &  16.45 &  16.39 &  16.38\\
&  0.046 &  0.034 &  0.025 &  0.026 &  0.016 &  0.018 &  0.016 &  0.018 &  0.018 &  0.017 &  0.025 &  0.021 &  0.026     \\
     U77 &  17.94 &  17.51 &  17.38 &  17.20 &  17.05 &  16.97 &  16.77 &  16.75 &  16.68 &  16.59 &  16.58 &  16.49 &  16.41\\
 & 0.066 &  0.047 &  0.042 &  0.039 &  0.025 &  0.024 &  0.033 &  0.024 &  0.030 &  0.028 &  0.030 &  0.031 &  0.044     \\
     H38 &  18.03 &  17.68 &  17.52 &  17.27 &  17.06 &  16.99 &  16.80 &  16.71 &  16.69 &  16.61 &  16.55 &  16.46 &  16.49\\
 & 0.035 &  0.028 &  0.021 &  0.020 &  0.016 &  0.016 &  0.015 &  0.015 &  0.016 &  0.015 &  0.022 &  0.019 &  0.026     \\
     C20 &  18.44 &  18.05 &  17.87 &  17.70 &  17.50 &  17.42 &  17.24 &  17.26 &  17.13 &  17.00 &  16.90 &  16.84 &  16.81\\
 & 0.037 &  0.026 &  0.022 &  0.022 &  0.017 &  0.020 &  0.017 &  0.020 &  0.019 &  0.019 &  0.025 &  0.021 &  0.039     \\
     C38 &  18.71 &  18.35 &  18.28 &  18.05 &  17.98 &  17.88 &  17.78 &  17.70 &  17.73 &  17.73 &  17.80 &  17.80 &  17.72\\
 & 0.080 &  0.050 &  0.040 &  0.058 &  0.029 &  0.040 &  0.032 &  0.037 &  0.034 &  0.041 &  0.058 &  0.061 &  0.089     \\
     H10 &  19.27 &  18.73 &  18.54 &  18.21 &  17.96 &  17.76 &  17.53 &  17.51 &  17.31 &  17.21 &  17.19 &  17.10 &  17.03\\
 & 0.093 &  0.055 &  0.043 &  0.038 &  0.030 &  0.030 &  0.033 &  0.031 &  0.029 &  0.029 &  0.033 &  0.036 &  0.053     \\
     U137&  19.21 &  18.88 &  18.68 &  18.42 &  18.22 &  18.10 &  17.88 &  17.80 &  17.81 &  17.74 &  17.71 &  17.54 &  17.61\\
 & 0.082 &  0.058 &  0.043 &  0.043 &  0.030 &  0.030 &  0.029 &  0.031 &  0.033 &  0.036 &  0.049 &  0.042 &  0.065     \\
   CBF11 & 19.56  & 19.05  & 18.99  & 18.82  & 18.50  & 18.52  & 18.19  & 18.11  & 17.99  & 17.95  & 17.67  & 17.62  & 17.54 \\
 & 0.153 &  0.098 &  0.080 &  0.085 &  0.063 &  0.075 &  0.052 &  0.053 &  0.050 &  0.067 &  0.081 &  0.046 &  0.071        \\
   CBF20 & 19.11  & 18.76  & 18.70  & 18.45  & 18.31  & 18.25  & 18.05  & 17.83  & 17.70  & 17.71  & 17.52  & 17.39  & 17.30 \\
 & 0.102 &  0.077 &  0.063 &  0.063 &  0.054 &  0.061 &  0.048 &  0.043 &  0.040 &  0.055 &  0.072 &  0.038 &  0.058        \\
   CBF22 & 18.66  & 18.23  & 18.13  & 18.01  & 17.76  & 17.79  & 17.74  & 17.63  & 17.60  & 17.49  & 17.56  & 17.49  & 17.32 \\
 & 0.065 &  0.046 &  0.037 &  0.040 &  0.032 &  0.038 &  0.035 &  0.034 &  0.035 &  0.043 &  0.071 &  0.040 &  0.056        \\
   CBF28 & 17.26  & 16.75  & 16.60  & 16.46  & 16.16  & 16.15  & 15.97  & 15.91  & 15.87  & 15.78  & 15.72  & 15.69  & 15.66 \\
 & 0.024 &  0.017 &  0.013 &  0.014 &  0.010 &  0.012 &  0.010 &  0.010 &  0.010 &  0.011 &  0.015 &  0.010 &  0.014        \\
   CBF49 & 19.37  & 18.54  & 18.56  & 18.52  & 18.08  & 18.07  & 17.86  & 17.91  & 17.93  & 17.74  & 17.80  & 17.81  & 17.75 \\
 & 0.151 &  0.073 &  0.065 &  0.078 &  0.049 &  0.057 &  0.046 &  0.050 &  0.054 &  0.059 &  0.094 &  0.059 &  0.090        \\
   CBF59 & 18.47  & 18.29  & 18.28  & 18.19  & 17.92  & 17.98  & 17.87  & 17.69  & 17.48  & 17.38  & 17.34  & 17.16  & 17.05 \\
 & 0.046 &  0.038 &  0.034 &  0.037 &  0.033 &  0.036 &  0.031 &  0.029 &  0.026 &  0.034 &  0.050 &  0.028 &  0.041        \\
   CBF69 & 19.36  & 19.10  & 18.99  & 18.53  & 18.62  & 18.48  & 18.34  & 18.28  & 18.19  & 18.21  & 17.96  & 17.90  & 18.08 \\
   & 0.154 &  0.125 &  0.135 &  0.155 &  0.100 &  0.088 &  0.093 &  0.087 &  0.099 &  0.103 &  0.106 &  0.102 &  0.163 \\
   CBF74 & 19.93  & 19.43  & 19.14  & 18.87  & 18.68  & 18.59  & 18.47  & 18.40  & 18.41  & 18.16  & 18.28  & 18.09  & 18.08 \\
   & 0.201 &  0.142 &  0.095 &  0.100 &  0.064 &  0.069 &  0.064 &  0.068 &  0.072 &  0.061 &  0.098 &  0.076 &  0.101 \\
   CBF77 & 18.53  & 18.33  & 18.35  & 18.04  & 18.08  & 17.80  & 17.82  & 17.82  & 17.64  & 17.50  & 17.57  & 17.46  & 17.09 \\
   & 0.061 &  0.045 &  0.039 &  0.046 &  0.032 &  0.036 &  0.030 &  0.036 &  0.035 &  0.028 &  0.049 &  0.037 &  0.054 \\
   CBF87 & 19.91  & 19.44  & 19.22  & 18.96  & 18.89  & 18.68  & 18.53  & 18.45  & 18.50  & 18.35  & 18.19  & 18.23  & 18.28 \\
   & 0.143 &  0.090 &  0.060 &  0.057 &  0.045 &  0.051 &  0.051 &  0.052 &  0.055 &  0.047 &  0.072 &  0.075 &  0.102 \\
\hline
\end{tabular}
\end{table}

\clearpage
\setcounter{table}{2}
\begin{table}[ht]
\caption{Continued}
\vspace {0.3cm}
\begin{tabular}{cccccccccccccc}
\hline
\hline
$\rm {Cluste}^{a}$ & 03  &  04 &  05 &  06 &  07 &  08 &  09 &  10 &  11 &  12 &  13 &  14 &  15\\
(1)    & (2) & (3) & (4) & (5) & (6) & (7) & (8) & (9) & (10) & (11) & (12) & (13) & (14)\\
\hline
   CBF87 & 19.91  & 19.44  & 19.22  & 18.96  & 18.89  & 18.68  & 18.53  & 18.45  & 18.50  & 18.35  & 18.19  & 18.23  & 18.28 \\
   & 0.143 &  0.090 &  0.060 &  0.057 &  0.045 &  0.051 &  0.051 &  0.052 &  0.055 &  0.047 &  0.072 &  0.075 &  0.102 \\
   CBF97 & 19.18  & 18.81  & 18.64  & 18.49  & 18.29  & 18.17  & 17.83  & 17.79  & 17.72  & 17.61  & 17.53  & 17.37  & 17.44 \\
   & 0.125 &  0.105 &  0.091 &  0.086 &  0.067 &  0.068 &  0.063 &  0.059 &  0.069 &  0.057 &  0.065 &  0.056 &  0.094 \\
  CBF112 & 19.08  & 18.79  & 18.74  & 18.51  & 18.49  & 18.40  & 18.33  & 18.33  & 18.22  & 18.04  & 18.08  & 18.07  & 18.17 \\
   & 0.082 &  0.071 &  0.059 &  0.063 &  0.053 &  0.056 &  0.059 &  0.065 &  0.071 &  0.063 &  0.080 &  0.082 &  0.161 \\
  CBF118 & 18.32  & 18.00  & 17.88  & 17.66  & 17.52  & 17.39  & 17.27  & 17.13  & 17.05  & 17.10  & 16.97  & 16.72  & 16.78 \\
   & 0.163 &  0.154 &  0.123 &  0.126 &  0.091 &  0.094 &  0.087 &  0.084 &  0.084 &  0.084 &  0.089 &  0.066 &  0.089 \\
  CBF119 & 18.48  & 18.16  & 18.09  & 17.87  & 17.72  & 17.60  & 17.54  & 17.34  & 17.26  & 17.31  & 17.16  & 16.90  & 16.95 \\
   & 0.160 &  0.154 &  0.122 &  0.126 &  0.093 &  0.094 &  0.096 &  0.082 &  0.086 &  0.081 &  0.086 &  0.063 &  0.089 \\
  CBF130 & 17.67  & 17.32  & 17.48  & 17.22  & 17.43  & 17.06  & 16.85  & 16.81  & 16.78  & 17.13  & 16.66  & 16.68  & 16.60 \\
   & 0.089 &  0.078 &  0.079 &  0.089 &  0.087 &  0.080 &  0.066 &  0.075 &  0.086 &  0.114 &  0.087 &  0.093 &  0.108 \\
  CBF131 & 18.27  & 17.87  & 17.87  & 17.75  & 17.54  & 17.48  & 17.32  & 17.32  & 17.32  & 17.34  & 17.21  & 17.24  & 17.08 \\
   & 0.113 &  0.089 &  0.074 &  0.079 &  0.052 &  0.054 &  0.049 &  0.052 &  0.061 &  0.058 &  0.069 &  0.077 &  0.085 \\
  CBF161 & 19.364 & 19.028 & 18.869 & 18.660 & 18.457 & 18.339 & 18.180 & 18.074 & 17.956 & 17.942 & 17.951 & 17.631 & 17.631\\
   & 0.077 &  0.054 &  0.039 &  0.034 &  0.033 &  0.033 &  0.034 &  0.036 &  0.035 &  0.038 &  0.059 &  0.043 &  0.078 \\
      M5 & 19.17  & 18.70  & 18.52  & 18.28  & 18.05  & 17.94  & 17.78  & 17.63  & 17.63  & 17.53  & 17.61  & 17.38  & 17.42 \\
         & 0.084 &  0.051 &  0.037 &  0.043 &  0.031 &  0.034 &  0.032 &  0.036 &  0.037 &  0.035 &  0.050 &  0.038 &  0.076 \\
     M12 & 17.99  & 17.67  & 17.53  & 17.29  & 17.15  & 17.04  & 16.83  & 16.78  & 16.74  & 16.64  & 16.53  & 16.41  & 16.36 \\
         & 0.116 &  0.114 &  0.089 &  0.086 &  0.064 &  0.064 &  0.038 &  0.051 &  0.055 &  0.051 &  0.056 &  0.058 &  0.061 \\
     M33 & 17.06  & 16.65  & 16.52  & 16.35  & 16.16  & 16.11  & 15.97  & 15.93  & 15.87  & 15.74  & 15.69  & 15.64  & 15.53 \\
         & 0.045 &  0.034 &  0.026 &  0.031 &  0.022 &  0.024 &  0.022 &  0.022 &  0.023 &  0.020 &  0.022 &  0.021 &  0.024 \\
     M51 & 18.07  & 17.70  & 17.64  & 17.52  & 17.35  & 17.32  & 17.21  & 17.18  & 17.12  & 17.06  & 17.01  & 16.98  & 16.85 \\
         & 0.030 &  0.022 &  0.018 &  0.021 &  0.017 &  0.019 &  0.019 &  0.022 &  0.023 &  0.024 &  0.038 &  0.035 &  0.042 \\
\hline
\end{tabular}
\end{table}
\vspace{-0.5cm}
$^{\rm a}~${CBF identifications are from Chandar, Bianchi, \& Ford (1999a, 2001);
M identifications are from Mochejska et al. (1998);
The others are from Christian \& Schommer (1982).}

\clearpage
\setcounter{table}{3}
\begin{table}[htb]
\caption[]{Metallicities of 31 old star clusters in M33}
\vspace {0.3cm}
\begin{tabular}{ccccc}
\hline
\hline
 Cluster & $B-V$ & $V-I$ & $\rm{[Fe/H]}$ & $\log$ age (yr) \\
\hline
   U49  & 0.68 & 1.029 & $-1.75\pm 0.036$& 9.60\\
   R12  & 1.03 & 1.154 & $-1.46\pm 0.036$& 10.00\\
   R14  & 0.98 & 1.311 & $-0.63\pm 0.036$& 9.11\\
   M9   & 0.69 & 1.016 & $-1.17\pm 0.108$& 9.63\\
   U77  & 0.67 & 0.994 & $-0.76\pm 0.048$& 9.20\\
   H38  & 0.73 & 1.070 & $-0.84\pm 0.036$&   9.70\\
   C20  & 0.77 & 1.045 & $-1.30\pm 0.072$&   9.95\\
   C38  & 0.73 & 0.883 & $-2.12\pm 0.108$&   9.28\\
   H10  & 0.96 & 1.243 & $-0.74\pm 0.036$&   9.90\\
   U137 & 0.83 & 1.099 & $-0.20\pm 0.072$&  10.27\\
  CBF11 & ...  & ...   & $-1.30\pm 0.240$&  10.30\\
  CBF20 & ...  & ...   & $-0.44\pm 0.240$&   9.54\\
  CBF22 & 0.513 & ...   & $-1.03\pm 0.481$&   9.26\\
  CBF28 & 0.794 & ...  & $-1.94\pm 0.240$&   9.80 \\
  CBF49 & 0.824 & ...  & $-1.69\pm 0.481$&   9.34 \\
  CBF59 & ...   & ... & $-0.14\pm 0.120$&   9.11\\ 
  CBF69 & ... & 1.061 &$-1.25\pm 0.120$&   9.76\\
  CBF74 & ... & 1.061 &$-0.12\pm 0.120$&   9.32\\
  CBF77 & ... & 0.438 &$-0.62\pm 0.240$&  9.01\\
  CBF87 & ... & 1.151 &$-0.97\pm 0.120$& 10.06\\
  CBF97 & ... & 1.015 &$-0.24\pm 0.120$& 10.28\\
  CBF112& ... & 0.846 &  $-2.98\pm 0.240$&  9.21\\
  CBF118& ... & 0.983 &  $0.13 \pm 0.120$&  9.16\\
  CBF119& ... & 0.940 &  $0.17 \pm 0.120$&  9.16\\
  CBF130& ... & 0.667 &  $-1.63\pm 0.240$&  9.06\\
  CBF131& ... & 0.898 &  $-1.19\pm 0.240$&  9.30\\
  CBF161& ... & ...   &  $0.32 \pm 0.120$& 10.28\\
  M5    & 0.73& 1.32  & $0.06 \pm 0.120$& 10.24\\
  M12  & 0.66& 1.11  & $-0.54\pm 0.120$&  9.63\\
  M33   & 0.71& 1.00  & $-1.51\pm 0.120$&  9.21\\
  M51   & 0.64& 0.65  &$-1.41\pm 0.120$&  9.01\\
\hline
\end{tabular}
\end{table}

\clearpage
\setcounter{table}{4}
\begin{table}[htb]
\caption[]{Comparison metallicity estimates with previous measurements}
\vspace {0.3cm}
\begin{tabular}{cccccc}
\hline
\hline
 Cluster & $\rm{{[Fe/H]}^{a}}$ & $\rm{{[Fe/H]}_{CPS}^{b}}$ & $\rm{{[Fe/H]}_{CS}^{c}}$ & $\rm{{[Fe/H]}_{BH}^{d}}$ & $\rm{{[Fe/H]}_{S}^{e}}$ \\
\hline
   U49  & $-1.75\pm 0.036$ & $-1.4$  & $-0.8\pm 0.3$ & $-1.70\pm 0.53$ & $-1.64\pm 0.20$ \\
   R12  & $-1.46\pm 0.036$ &  ...    & $-1.2\pm 0.3$ &     ...         & $-1.19\pm 0.24$ \\
   R14  & $-0.63\pm 0.036$ &  ...    & $-1.5\pm 0.3$ &     ...         & $-1.00\pm 0.50$ \\
   M9   & $-1.17\pm 0.108$ & $-2.2$  & $-1.7\pm 0.3$ &     ...         & $-1.64\pm 0.28$ \\
   U77  & $-0.76\pm 0.048$ &  ...    &     ...       & $-1.77\pm 0.77$ & $-1.56\pm 0.30$ \\
   H38  & $-0.84\pm 0.036$ & $-1.0$  & $-1.5\pm 0.3$ &     ...         & $-1.10\pm 0.10$ \\
   C20  & $-1.30\pm 0.072$ & $-1.1$  & $-2.2\pm 0.3$ & $-1.25\pm 0.79$ & $-1.25\pm 0.22$ \\
   C38  & $-2.12\pm 0.108$ &  ...    & $-1.2\pm 0.3$ &     ...         & $-0.65\pm 0.16$ \\
   H10  & $-0.74\pm 0.036$ &  ...    &     ...       & $-0.91\pm 0.90$ & $-1.44\pm 0.26$ \\
   U137 & $-0.20\pm 0.072$ &  ...    &     ...       & $-0.12\pm 0.38$ & $-0.98\pm 0.16$ \\
\hline
\end{tabular}\\
\end{table}
\vspace{-0.5cm}
$^{\rm a}~${This paper}

$^{\rm b}~${Cohen, Persson, \& Searle 1984}

$^{\rm c}~${Christian \& Schommer 1988}

$^{\rm d}~${Brodie \& Huchra 1991}

$^{\rm e}~${Sarajedini et al. 1998}

\end{document}